\documentclass[reqno,12pt]{amsart}

\usepackage{graphicx,epsfig,rotating}
\usepackage[centertags]{amsmath}
\usepackage{amsfonts}
\usepackage{amssymb,amsmath}
\usepackage{amsxtra}
\usepackage[active]{srcltx}
\usepackage{dcolumn}

\newcommand{\dif}{\partial}

\newcommand{\leqs}{\leqslant}

\newcommand{\ev}[1]{\langle #1 \rangle}

\newcommand{\occ}{{\mathrm{oc}}}

\DeclareMathOperator{\const}{const}
\DeclareMathOperator{\sign}{sign}

\newcommand{\al}{\alpha}     \newcommand{\be}{\beta}
     \newcommand{\de}{\delta}
   \newcommand{\ve}{\varepsilon}

\newcommand{\la}{\lambda}    \newcommand{\om}{\omega}
    
  \newcommand{\te}{\theta}
\newcommand{\vt}{\vartheta}  

\newcommand{\De}{\Delta}

\begin{document}

\title[Collective surface diffusion near a phase transition]{Collective surface diffusion near a first-order phase transition}

\author[I.~Medve\v d and A.~Trn\'{\i}k]{Igor Medve\v d and Anton Trn\'{\i}k}
\address{Department of Physics, Constantine the Philosopher University, 949 74 Nitra, Slovakia}
\address{Department of Materials Engineering and Chemistry, Czech Technical University, 16629 Prague, Czech Republic}%
\email{imedved@ukf.sk}%

%


\keywords{Surface diffusion; thermodynamic factor; coverage; susceptibility.}

\begin{abstract}
For a large class of lattice models we study the thermodynamic factor, $\Phi$, of the collective surface diffusion coefficient near a first-order phase transition between two phases at low temperatures. In a two-phase regime its dependence on the coverage, $\te$, is $\Phi \approx \te/[(\te-\te_-)(\te_+ - \te)N]$, where $N$ is the number of adsorption sites and $\te_\pm$ are the single-phase coverages at the transition. In the crossover between the two-phase and single-phase regimes $\Phi(\te)$ is shown to have a more complex behavior. The results are applied to a simple $2D$ lattice model.
\end{abstract}


\maketitle



Surface diffusion is an intriguing phenomenon that has attracted much attention of surface scientists in the past few decades \cite{Ala02,Nau05}. One of the relevant transport coefficients for surface diffusion is the collective (or chemical) surface diffusion coefficient, $D_c$. It is associated with the decay of fluctuations in the adparticle density at large time and space intervals and is  defined via the Fick's first law, $\mathbf{J} = - D_c \nabla \te$, where $\mathbf{J}$ is the surface diffusion flux and $\te$ is the surface coverage.

Considering suitable $2D$ lattice gas models, computer simulation methods have been frequently applied in theoretical studies of $D_c$ (and of surface diffusion in general) with a particular interest in the presence of phase transitions and their effects on surface diffusion \cite{UG95,Zh95,Ta01,Ta03,Za04,Chv06,Za07,Chv08,Ta08,Za10}. Usually, the observation that $D_c = \Phi \, D_{\mathrm{CM}}$ \cite{Ala02} was used, where $D_{\mathrm{CM}}$ is the center of mass diffusion coefficient and represents kinetic properties of $D_c$, while $\Phi = \ev{N_\occ}/(\ev{N_\occ^2} - \ev{N_\occ}^2)$ is associated with thermodynamic properties of $D_c$ and is called the thermodynamic factor ($N_\occ$ is the number of adsorption sites in a system occupied by adparticles and $\ev{ \cdot }$ denotes the statistical mean value). The factor is simply related to the coverage $\te = \ev{N_\occ}/N$ and isothermal susceptibility $\chi = \be (\ev{N_\occ^2} - \ev{N_\occ}^2)/N$,
\begin{equation}\label{eq: Phi}
  \Phi = \be \, \frac{\te}{\chi} \,,
\end{equation}
where $N$ is the total number of adsorption sites in the system and $\be = 1/k_B T$ is the inverse temperature. In the hydrodynamic limit (when $\te$ varies slowly with space and time) $D_c$ can be obtained from the system free energy, and so it is a purely thermodynamic quantity \cite{Zh85,Zh91,Ta98}.

In this letter we use a rigorous statistical mechanical approach to study $\Phi$ \emph{near} a first-order phase transition with a (in general asymmetric) coexistence of two phases (denoted as `$+$' and `$-$'). To this end, we consider a finite array of $N$ adsorption sites with periodic boundary conditions and use that, for a wide range of classical lattice spin models (such as the models with a finite range $m$-potential and a finite number of ground states), the coverage and susceptibility in the array have a remarkably universal behavior near such a transition \cite{BoKo90,BoKo92}. Hence, the behavior of $\Phi$ near the transition as a function of the chemical potential, $\mu$, as well as of the coverage $\te$ can be described in a very general and unifying way. It is only needed that the temperature is low ($\be$ is large) and that the array is large and not too oblong ($N$ is large and the number of boundary sites in the array is of order of a fractional power of $N$).

\begin{figure}
  \centering
  \includegraphics{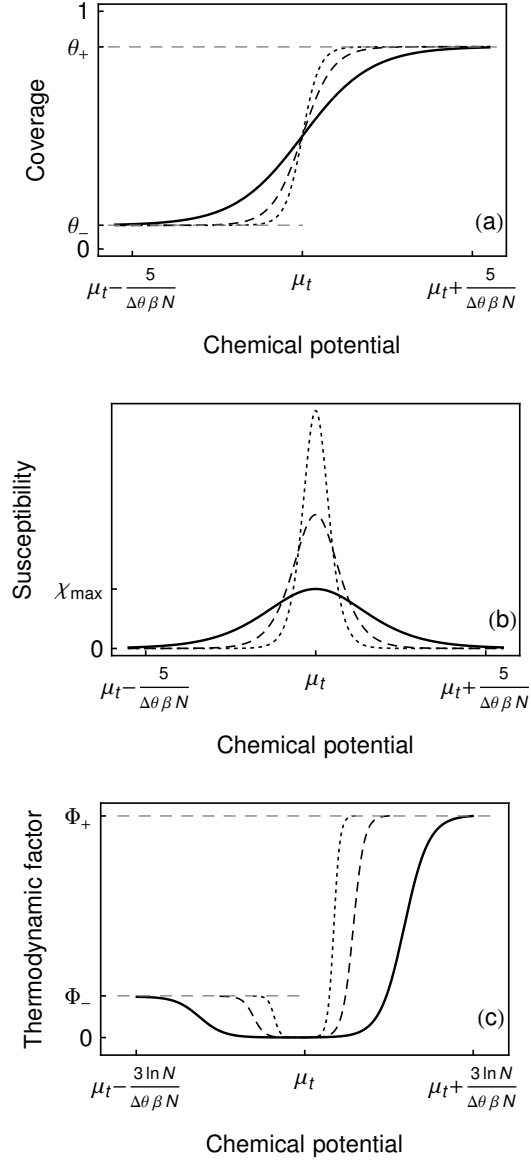}
  \caption{The dependencies of (a) coverage, (b) susceptibility, and (c) thermodynamic factor on the chemical potential near the transition (full lines). For two larger sizes of a system the dependencies are depicted by dashed and dotted curves. The points $\chi_{\max} \approx (\De\te/2)^2 \be N$ and $\Phi_\pm \approx \be \te / \chi_\pm$.}
  \label{fig: on mu}
\end{figure}

\emph{The $\mu$-dependence of $\Phi$ near the transition.} Generally, the coverage can be expressed as \cite{BoKo90}
\begin{equation} \label{eq: te per a}
  \te = (-f'_+) \la_+ + (-f'_-) \la_- + R_N,
\end{equation}
where $f_\pm$ and $-f'_\pm$ are single-phase specific free energies and coverages, respectively, the weight factors $\la_\pm = 1/\{1 + \exp[\pm \be (f_+ - f_-) N]\}$, and the error term $R_N = O(\exp(-c \be \sqrt N))$ with $c>0$ (primes denote derivatives with respect to $\mu$, and $O(x)$ stands for a term that can be bounded by $\const x$). To get an explicit $\mu$-dependence of $\te$ near a transition point, $\mu_t$, one applies Taylor expansions around $\mu_t$, yielding \cite{BoKo90}
\begin{equation} \label{eq: te per b}
  \te = \frac{\vt_0}2 + \frac{\De\vt}2 \tanh \frac Y2
  + O(\be^2 |\mu - \mu_t|^2) + R_N
\end{equation}
(see Fig.~\ref{fig: on mu}(a)). The shorthands $q_0 = q_+ + q_-$ and $\De q = q_+ - q_-$ denote the sum and difference, respectively, of any single-phase quantities $q_\pm$, $\vt_\pm = \te_\pm + \chi_\pm (\mu - \mu_t)$, and $Y = \be N (\mu - \mu_t) [ \De\te + \De\chi \, (\mu - \mu_t)/2 ]$, where $\te_\pm = -f'_\pm (\mu_t)$ and $\chi_\pm = -f''_\pm (\mu_t)$ are the single-phase coverages and susceptibilities at the transition. If the plus (minus) corresponds to the phase stable above (below) $\mu_t$, the coverage jump at $\mu_t$ is $\te(\mu_t+0) - \te(\mu_t-0) = \De\te > 0$. For $\mu$ beyond the transition Eq.~\eqref{eq: te per a} yields another expression for $\te$ \cite{BoKo90},
\begin{equation} \label{eq: te inf}
  \te = \te_\infty + P_N + R_N.
\end{equation}
The infinite-volume coverage $\te_\infty$ is equal to $-f'_+$ ($-f'_-$) for $\mu$ above (below) $\mu_t$ and $P_N = O(\exp(-C |\mu-\mu_t| \be N))$ with $C>0$. Using $s = \sign (\mu-\mu_t)$ both for numbers $\pm 1$ and for subscripts `$\pm$', we may write $\te_\infty = -f'_s$.

Equation~\eqref{eq: te per a} holds also for the $\mu$-derivatives of $\te$ with the same error term $R_N$ \cite{BoKo90}. So, we can readily obtain a general expression for the susceptibility,
\begin{equation} \label{eq: chi per a}
  \chi = (-\De f')^2 \be N \la_+ \la_-
  + (-f''_+) \la_+ + (-f''_-) \la_- + R_N,
\end{equation}
where $-f''_\pm$ are the single-phase susceptibilities. The $\mu$-dependence of $\chi$ then follows analogously to that of $\te$: either from Taylor expansions around $\mu_t$ (useful near the transition),
\begin{align} \label{eq: chi per b}
  \chi &= \frac{\De\vt^2 \be N}{4 \cosh^2 \frac Y2}
  + \frac{\chi_0}2 + \frac{\De\chi}2 \tanh \frac{Y_1}2
\\
  &\quad + O(\be^3 N |\mu - \mu_t|^2)
  + O(\be^2 |\mu - \mu_t|) + R_N
\end{align}
with $Y_1 = \De\te \be N (\mu - \mu_t)$ (see Fig.~\ref{fig: on mu}(b)), or by using the infinite-volume susceptibility $\chi_\infty = -f''_s$ (useful beyond the transition),
\begin{equation}
   \label{eq: chi inf}
  \chi = \chi_\infty + O(\be N P_N) + R_N.
\end{equation}

These formulas for $\te$ and $\chi$ immediately yield the $\mu$-dependence of the thermodynamic factor $\Phi$. Near the transition region, $|\mu - \mu_t| \ll 1$, it is given by Eqs.~\eqref{eq: te per b} and \eqref{eq: chi per b} (see Fig.~\ref{fig: on mu}(c)), while farther away from the transition, $|\mu - \mu_t| \gg 1/N$, it is given by Eqs.~\eqref{eq: te inf} and \eqref{eq: chi inf}.

At a certain distance from $\mu_t$ (both below and above $\mu_t$) there are two narrow regions of $\mu$ where $\te$ and $\chi$ reduce to their infinite-volume limits $\te_\infty$ and $\chi_\infty$ so that the behavior of $\Phi$ switches from a two-phase regime to one of the two single-phase regimes (see Fig.~\ref{fig: on mu}(c)). According to Eqs.~\eqref{eq: te per b} and \eqref{eq: chi per b}, these two crossover regions occur when $\be N \exp(-|Y|)$ becomes small, i.e., within the intervals $\de\mu_- \leqs |\mu - \mu_t| \leqs \de\mu_+$, where $\de\mu_\pm = (\ln N \pm K\ln\ln N)/\De\te \be N$ with $K > 0$ large.

\begin{figure}
  \centering
  \includegraphics{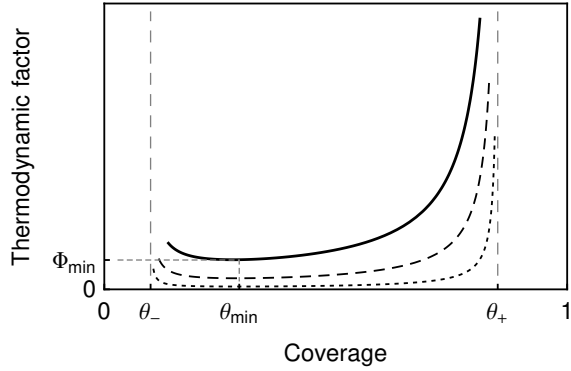}
  \caption{The $\te$-dependence of $\Phi$ (the full line) in the interval $t_- \leqs \te \leqs t_+$ (the two-phase regime) as given by Eq.~\eqref{eq: Phi 1 per}. For two larger sizes of a system the dependence is depicted by a dashed and a dotted curve. A minimum of $\Phi$ is attained at $\te_{\min} \approx (\te_- \te_+)^{1/2}$ and its value is $\Phi_{\min} \approx 1/(\sqrt{\te_-} - \sqrt{\te_+})^2 N$.}
  \label{fig: PhiTwoPh}
\end{figure}

\emph{The $\te$-dependence of $\Phi$ near the transition.} The coverage dependence of $\Phi$ follows upon the elimination of $\mu$ between $\te(\mu)$ and $\chi(\mu)$. This procedure is simple to achieve within the two-phase regime, $|\mu - \mu_t| \leqs \de\mu_-$, because it reduces to the elimination of $Y$ between Eqs.~\eqref{eq: te per b} and \eqref{eq: chi per b}. We readily get
\begin{equation} \label{eq: Phi 1 per}
  \Phi = \frac\te{(\te - \te_-) (\te_+ - \te) N + O(\ln N)} \,,
  \quad
  t_- \leqs \te \leqs t_+,
\end{equation}
where $t_\pm = \te(\mu_t \pm \de\mu_-)$ are the most extreme values that $\te$ attains in the two-phase regime. Since $t_\pm = \te_\pm \mp \De\te \ln^K N/N + O(\ln N/N)$ by Eq.~\eqref{eq: te per b}, we have $\te_- < t_- < t_+ < \te_+$. The behavior of $\Phi$ as given by Eq.~\eqref{eq: Phi 1 per} is shown in Fig.~\ref{fig: PhiTwoPh}.

Next, consider the range $\de\mu_- \leqs |\mu - \mu_t| \leqs d\mu$ with $d\mu = 1/\De\te \be N^{3/4}$ that contains both crossover regions as well as a part of either single-phase regime (instead of $N^{3/4}$ any power $N^\al$ with $1/2 < \al < 1$ can be considered). To eliminate $\mu$ between $\te(\mu)$ and $\chi(\mu)$ we rewrite Eqs.~\eqref{eq: te per b} and \eqref{eq: chi per b} as $\te = \te_s + s \chi_s |\mu-\mu_t| - s \De\te A + Q_N$ and $\chi = \De\te^2 \be N A + \chi_s + \be N Q_N$, respectively, with $A = \exp(-|Y_1|)$ and $Q_N = O(\ln^K N/N^{3/2})$. Eliminating $A$ between these equations, we get
\begin{equation} \label{eq: Phi 2 per}
  \Phi =
  \left\{
  \begin{array}{ll}
  \frac\te{ \frac{\chi_-}\be \, (\om_- + 1) + O (N Q_N)},
    & \quad \tau_- \leqs \te \leqs t_-, \\[3mm]
  \frac\te{ \frac{\chi_+}\be \, (\om_+ + 1) + O (N Q_N)},
    & \quad t_+ \leqs \te \leqs \tau_+,
  \end{array}
  \right.
\end{equation}
with
\begin{equation*}
  \om_\pm = W \Bigl( \frac{\De\te^2 \be N}{\chi_\pm} \,
  \exp \Bigl[
    \pm \frac{\De\te \be N}{\chi_\pm} \, (\te_\pm - \te)
  \Bigr] \Bigr),
\end{equation*}
where $W(y)$ is the Lambert $W$-function (i.e., the inverse to $y = W \exp W$) and the restrictions on $\te$ in Eq.~\eqref{eq: Phi 2 per} correspond to the two intervals of values attained by $\te$ in the considered $\mu$-region below and above $\mu_t$. Thus, $\tau_\pm = \te(\mu_t \pm d\mu) = \te_\pm \pm \chi_\pm / \De\te \be N^{3/4} + O(1/N^{3/2})$ by Eq.~\eqref{eq: te per b} so that $\tau_- < \te_- < t_- < t_+ < \te_+ < \tau_+$.

\begin{figure}
  \centering
  \includegraphics{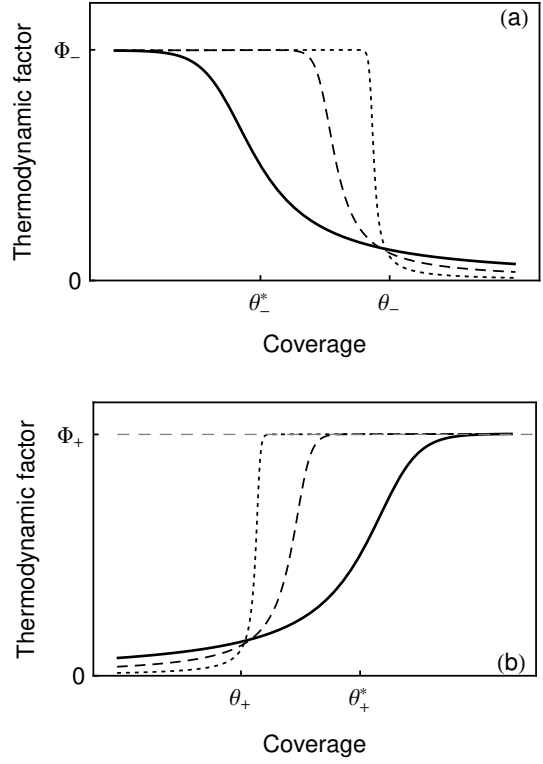}
  \caption{The $\te$-dependence of $\Phi$ (full lines) in the crossover regions (a) near $\te_-$ and (b) near $\te_+$ as given by Eq.~\eqref{eq: Phi 2 per} (the error term $O(N Q_N)$ is neglected). For two larger sizes of a system the dependencies are depicted by dashed and dotted curves.}
  \label{fig: PhiCross}
\end{figure}

The behavior of $\Phi$ as given by Eq.~\eqref{eq: Phi 2 per} is shown in Fig.~\ref{fig: PhiCross}. If $\om_\pm \gg 1$, the two-phase behavior of $\Phi$ prevails, while a single-phase behavior is dominant for $\om_\pm \ll 1$. The crossover occurs when $\om_\pm$ is comparable to $1$, i.e., when $\te$ is comparable to the value, $\te_\pm^* = \te_\pm \pm \, (\chi_\pm/\De\te \be N) \,
[ \ln (\De\te^2 \be N/\chi_\pm) - 1 ]$, at which $\om_\pm = 1$. For $\te \lesssim \te_-^*$ (for $\te \gtrsim \te_+^*$) the thermodynamic factor is increasing in approximately a linear way as $\be\te/\chi_-$ (as $\be\te/\chi_+$), while as $\te$ crosses $\te_-^*$ ($\te_+^*$) and gets above $\te_-$ (below $\te_+$) it approaches the two-phase behavior described by Eq.~\eqref{eq: Phi 1 per}.


%
%
\begin{figure}
  \centering
  \includegraphics{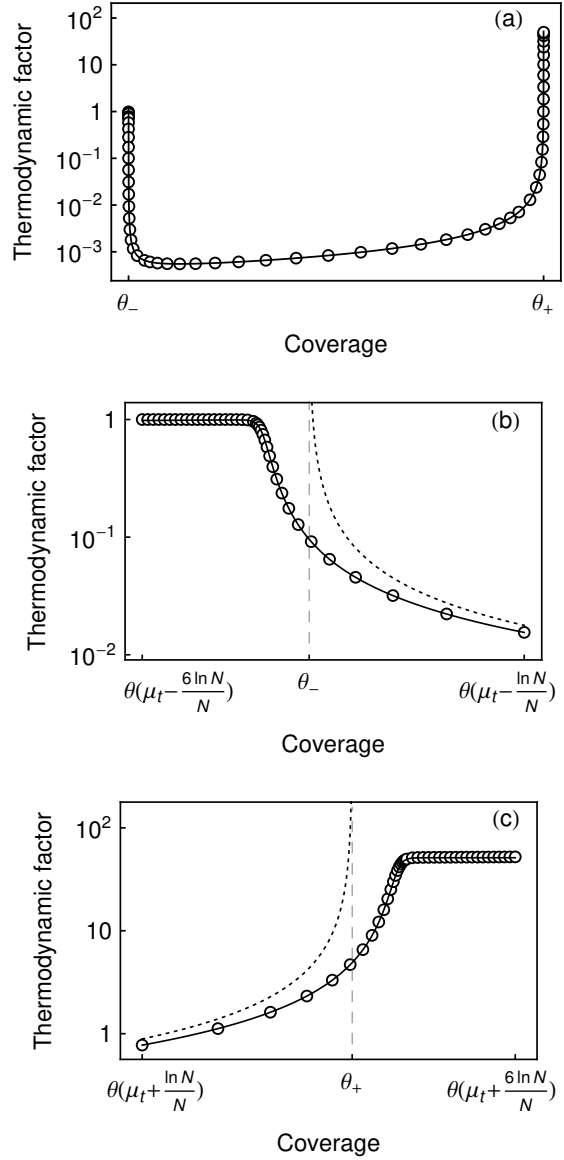}
  \caption{The $\te$-dependence of $\Phi$ for the example model with $N = 50 \times 50$ and $\be|\ve| = 1.2 \ln 3$ in (a) the two-phase regime, (b) the crossover region below $\mu_t$, and (c) the crossover region above $\mu_t$. Circles are values of $\Phi$ obtained numerically upon elimination of $\mu$ between $\te(\mu)$ and $\chi(\mu)$ using Eqs.~\eqref{eq: te per a} and \eqref{eq: chi per a}. Full lines correspond to formulas~\eqref{eq: Phi 1 per} (case (a)) and \eqref{eq: Phi 2 per} (cases (b) and (c)).  Dotted lines in (b) and (c) correspond to the two-phase regime formula~\eqref{eq: Phi 1 per}.}
  \label{fig: Ising}
\end{figure}

\emph{An example.} To illustrate the above results, we consider the following one-component lattice-gas model of diffusion on a triangular lattice \cite{UG95,Ta01,Ta03}. Each lattice site is either occupied by an adparticle or is vacant. The model Hamiltonian is $H = \ve N_\occ^b - \mu N_\occ$, where $N_\occ^b$ is the number of nearest-neighbor bonds occupied by adparticles and $\ve < 0$ is an attractive interaction between two closest adparticles. The model is equivalent to the standard Ising model, and at low temperatures (for $\be|\ve| > \ln 3$) it exhibits a first-order phase transition at $\mu_t = 3\ve$ between a fully vacant phase (stable below $\mu_t$ and thus corresponding to $-$) and a fully occupied phase (stable above $\mu_t$ and corresponding to $+$) \cite{HM02}.

As follows from Ref.~\cite{BoKo90}, one has $f_\pm = e_\pm + s_\pm$, where $e_+ = 3\ve-\mu$ and $e_- = 0$ are the specific ground state energies and $s_\pm$ represent thermal excitations over the ground states and can be expressed by cluster expansions. In the lowest order approximation $s_\pm \approx - \exp(-\be \De H_\pm)/\be$, where $\De H_+ = \mu - 6\ve$ and $\De H_- = -\mu$ are the energy excesses of the lowest excited states (one adparticle taken from and added to the corresponding ground state). So, $-f'_+ \approx 1 - \eta \exp[-\be (\mu-\mu_t)]$, $-f'_- \approx \eta \exp[\be (\mu-\mu_t)]$, and $-f''_\pm \approx \be \eta \exp[\mp \be (\mu-\mu_t)]$ with $\eta = \exp(\be\mu_t)$. The thermodynamic factor $\Phi$ of the model can be readily obtained from these expressions and Eqs.~\eqref{eq: te per a} and \eqref{eq: chi per a}, and it is shown in Fig.~\ref{fig: Ising}. Moreover, in the single-phase regimes, $|\mu-\mu_t| \geq d\mu$, Eqs.~\eqref{eq: te inf} and \eqref{eq: chi inf} yield $\Phi \approx 1 + O(N P_N)$ for $\mu$ below $\mu_t$ (i.e., for $\te$ near $0$) and $\Phi \approx \te/(1-\te) + O(N P_N)$ for $\mu$ above $\mu_t$ (i.e., $\te$ near $1$). Thus, $D_c \approx D_{\mathrm{CM}}$ at low coverages, while the Langmuir relation holds at high coverages.

Finally, $D_c = D_c^0 \be \exp(\be\mu) P_{00}/\chi$ in the hydrodynamic limit \cite{Ta01}, where $D_c^0$ is the diffusion coefficient of non-interacting adparticles and $P_{00} = 1 - 2\te + (\dif f/\dif \ve)/3$ is the probability that two nearest-neighbor sites are empty ($f = - k_B T \ln Z/N$ is the system specific free energy). Using the results of Ref.~\cite{BoKo90}, we get $\dif f/\dif \ve = [(\dif f_+/\dif \ve) \la_+ + (\dif f_-/\dif \ve) \la_-](1 + R_N)$. Along with the above formulas, we find that at $\mu = \mu_t$, say, one has $\ln(D_c/D_c^0) \approx -3|\ve|/k_B T - \ln [(1-2\eta)^2 N/2] + O(1/N)$.

This research was supported by the VEGA under Grant No.~1/0302/09 and by the Ministry of Education, Youth and Sports of the Czech Republic, under the project No.~MSM: 6840770031.



\end{document}